# Conditions determining the morphology and nanoscale magnetism of Co nanoparticles: Experimental and numerical studies

K. Simeonidis[1], C. Martinez-Boubeta[2], O. Iglesias[3], A. Cabot[2,4], M. Angelakeris[1], S. Mourdikoudis[1], I. Tsiaoussis[1], A. Delimitis[5], C. Dendrinou-Samara[6], O. Kalogirou[1]

[1] Department of Physics, Aristotle University of Thessaloniki, 54124 Thessaloniki, Greece.
[2] Departament d' Electrònica and Institute of Nanoscience and Nanotechnology (IN$^2$UB), Facultat de Física, Universitat de Barcelona, Av. Diagonal 647, 08028 Barcelona, Spain.
[3] Departament de Física Fonamental and Institute of Nanoscience and Nanotechnology (IN$^2$UB), Facultat de Física, Universitat de Barcelona, Av. Diagonal 647, 08028 Barcelona, Spain.
[4] Catalonia Institute for Energy Research – IREC, Jardins de les Dones de Negre 1, 08930, Sant Adrià del Besòs, Barcelona, Spain.
[5] Chemical Process Engineering Research Institute (CPERI), Centre for Research & Technology – Hellas (CERTH), 57001 Thessaloniki, Greece.
[6] Department of Chemistry, Aristotle University of Thessaloniki, 54124 Thessaloniki, Greece.

**Abstract**

Co-based nanostructures ranging from core-shell to hollow nanoparticles were produced by varying the reaction time and the chemical environment during the thermal decomposition of $Co_2(CO)_8$. Both structural characterization and kinetic model simulation illustrate that the diffusivities of Co and oxygen determine the growth ratio and the final morphology of the nanoparticles. Exchange coupling between Co and Co-oxide in core/shell nanoparticles induced a shift of field-cooled hysteresis loops that is proportional to the shell thickness, as verified by numerical studies. The increased nanocomplexity when going from core/shell to hollow particles, also leads to the appearance of hysteresis above 300 K due to an enhancement of the surface anisotropy resulting from the additional spin-disordered surfaces.

# I. INTRODUCTION

Cobalt-based nanoparticles reside among the most promising materials for technological applications like information storage, magnetic fluids and catalysis [1,2]. The low crystal anisotropy of cobalt also motivates their study as a model system for the effects of size, shape, crystal structure, and surface anisotropy on their macroscopic magnetic response. In order to produce monodisperse cobalt nanoparticles a variety of preparation methods have been reported [3]. High-temperature chemical methods like the reduction of a cobalt salt or the thermal decomposition of a cobalt carbonyl, resulting in colloid dispersions of the nanoparticles, are the most common, since they provide a good control over their geometrical and structural features. However, as-synthesized Co nanoparticles are susceptible to oxidation when exposed to air or even under inert environment [4]. Depending on the presence of oxidants and the particle size, Co NPs can be either fully oxidized to cobalt oxides or partially oxidized to form shells of native oxides on the particle surface. Usually, natural oxidation in cobalt nanoparticles larger than 5 nm is restricted to an outer shell, thus it is possible to maintain a stable cobalt core for a long period after synthesis.

In a Co/CoO interface, the proximity between a ferromagnetic (FM) and an antiferromagnetic (AFM) material leads not only to a structural modification, but also to a competition between different types of magnetic ordering. Particularly, the exchange coupling at a FM/AFM interface may induce unidirectional anisotropy in the FM, below the Néel temperature of the AFM, causing a shift in the hysteresis loop, a phenomenon known as exchange bias [5]. In recent years, the study of exchange bias in nanoparticles and nanostructures has gained renewed interest. It has been shown that the control of the core/shell interactions or exchange coupling between the particle surface and the embedding matrix can increase the superparamagnetic limit, providing an advantage for their use in magnetic recording media, permanent magnets and spintronics [6].

Although many numerical studies predicting the mechanism and the factors that determine the exchange bias value in FM/AFM nanoparticles were reported [7], further experimental investigation is required mainly concerning the size dependence of the oxidation extent, the critical diameter for the appearance of a stable, under ambient conditions, Co/CoO interface and the role of the shell thickness or even the actual interface geometry in the exchange coupling intensity. Moreover, the multi-crystalline structure of single-phase nanoparticles and their shape deviation from the typical spherical geometry could be a reason for the appearance of exchange coupling on fully oxidized nanoparticles due to the noncollinear spin configuration or spin canting at the particle surface. Recently, hollow ferrimagnetic nanoparticles ($\gamma$-$Fe_2O_3$) were investigated in order to correlate crystallographic and magnetic domain arrangement with surface effects [8]. Additionally, pure antiferromagnetic nanoparticles were reported to exhibit net magnetization arising from uncompensated surface spins on the surface [3]. Yet, another source of anisotropy in hollow nanoparticles formed through a mechanism similar to the Kirkendall effect is the presence of a large number of voids in the structure, a possible small core of Co standing in the center of the nanoparticle [9,10] or asymmetrically localized [11]. For instance, the formation of CoO hollow and Co/CoO yolk-shell nanoparticles is attributed to the condensation of pores produced by the different diffusion rates between cobalt and oxygen atoms [12,13]. In fact, simulations of the growth of hollow particles using a kinetic model for the particle oxidation showed that different particle geometries are achieved depending on the particle size and oxidation conditions [14,15]. In spite of these indications, not much attention was given in the study of the magnetic properties of such systems.

This work concentrates on the nanoscale magnetic behavior of different partially and fully oxidized Co nanoparticles in the size range 3-18 nm. Controlling the thermal decomposition method, various configurations of nanoparticles ranging from solid or hollow

Co-oxide to core-shell or yolk-shell Co/CoO were prepared. The appearance of exchange coupling when metal cobalt and cobalt oxide co-exist was evaluated so as to understand the critical role of geometrical and morphological parameters on the observed magnetic properties. With this aim, a kinetic model to explain the geometry evolution of Co nanocrystals during the oxidation reaction, and Monte Carlo (MC) computing to simulate the high degree of spin disorder present at the particle surfaces, were used.

## II. EXPERIMENTAL

The nanoparticles analyzed in this study were synthesized by the thermal decomposition of $Co_2(CO)_8$ at moderate temperatures (180 °C) using oleic acid and trioctylphosphine-oxide (TOPO) as capping agents. In all cases, diphenylether was used as solvent. Precursor-to-surfactant and oleic acid-to-TOPO molar ratios were both set to 2/1, in order to avoid the production of elongated particles and nanorods or the formation of stable complexes [16]. In a typical preparation, 1.8 mmol of $Co_2(CO)_8$ were added into a spherical flask containing a preheated mixture of oleic acid (0.6 mmol) and TOPO (0.3 mmol) in phenylether (20 ml) at 135 °C. Then, the temperature was increased and maintained at 180 °C for 90 min before cooling to room temperature. Both reaction time and temperature were observed to have a key role on the final nanocrystal size. The final product was centrifuged and washed with hexane to remove any non-reacted compounds and excessive solvent. Different nanoparticle oxidation protocols were used in order to obtain nanocrystalline heterostructures with distinct geometries: i) A set of nanoparticles was oxidized at room temperature in an ambient atmosphere to reproduce usual oxidation conditions; ii) A thermal oxidation process at 180 °C in a 20% oxygen atmosphere was employed for the oxidation of a second set of nanocrystals; iii) The most radical oxidation process involved the introduction of pyridine N-oxide in the reaction solution at a temperature equal or close to the reaction one. Pyridine N-oxide is a well known oxidizing agent that promotes Co oxidation [17].

The structural characterization of the samples was mainly performed by X-ray powder diffraction (XRD) on a Philips PW 1820 diffractometer with CuK$_a$ radiation. Low magnification TEM images of the prepared materials, acquired with a JEOL 120 CX microscope at a 100 kV acceleration voltage, were used to calculate the average size of the nanoparticles and provide a view of their morphology and spatial arrangement. Supplementary information about structure and composition down to the atomic level of the individual particles composition was provided by high-resolution TEM (HRTEM) observations using a JEOL 2011 microscope operating at 200 kV, with a point resolution of 0.23 nm. The magnetic properties of the nanoparticles were investigated using a MPMS-5 superconducting quantum interference device (SQUID) magnetometer at magnetic fields up to 5 T. Hysteresis loops were measured after field cooling at 20 kOe. The temperature-dependent magnetization curves were recorded after cooling the sample to 5 K either in a zero magnetic field (zero-field-cooled ZFC) or in a 100 Oe magnetic field (field-cooled FC).

## III. RESULTS AND DISCUSSION

### A. STRUCTURAL CHARACTERIZATION

Figure 1 shows representative TEM images exhibiting the geometrical characteristics of the six samples studied in the present work. An overview of the synthetic parameters and the resulting morphological and structural characteristics of the studied samples is given in Table 1.

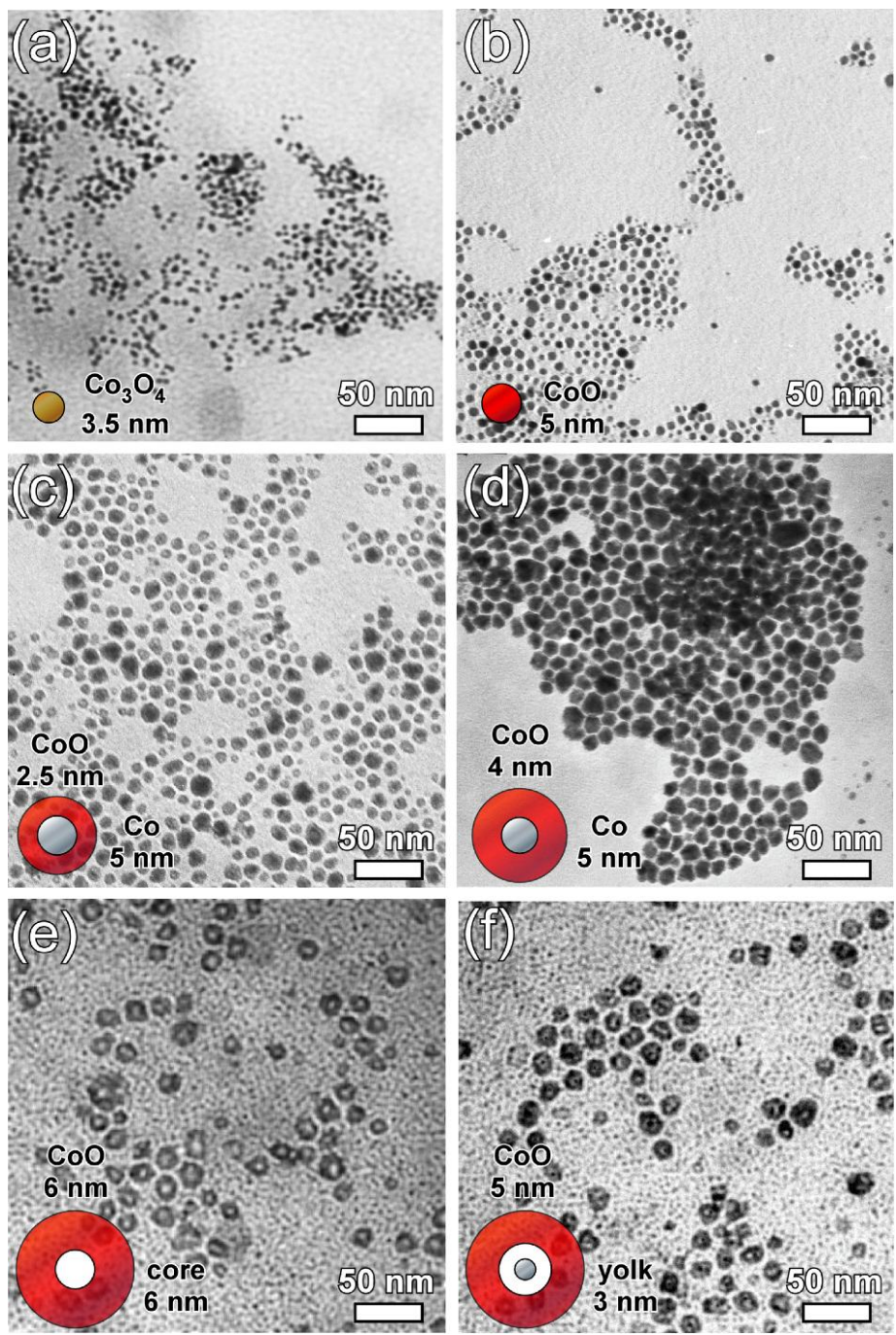

**FIG. 1.** TEM images of samples 1-6 (a)-(f).

Sample 1 was prepared by the decomposition of cobalt carbonyl in the presence of pyridine N-oxide. The addition of the oxidizing element in the initial solution strongly affects the growth mechanism and results in 3.5 nm completely oxidized solid nanoparticles [sample 1, Fig. 1(a)]. As shown by XRD patterns [Fig. 2(a)], pyridine-N-oxide favors the oxidation of cobalt to $Co_3O_4$, which is further facilitated by the simultaneous use of open-air atmosphere during the reaction, instead of inert one.

**Table 1.** Parameters of synthesis, morphological and structural characteristics of samples

| Sample | Conditions | Temperature (°C) | Reaction-aging (min) | Mean size (nm) | Composition | Magnetization* at 300 K ($Am^2/kg$) |
|---|---|---|---|---|---|---|
| 1 | Open air/ oxidizing | 180 | 45 | 3.5±0.3 | $Co_3O_4$ | 1.3 |
| 2 | Inert | 180 | 45 | 5.1±0.3 | CoO | 1.7 |
| 3 | Inert | 180 | 60 | 9.9±1.3 | Co/CoO | 24.9 |
| 4 | Inert | 180 | 90 | 13.0±1.2 | Co/CoO | 9.2 |
| 5 | Open air/ oxidizing | 180–150 | 90–120 | 18.1±1.8 | CoO | 2.1 |
| 6 | Open air | 180 | 90 | 17.8±1.7 | Co/CoO | 1.9 |

*measured at 2 T

When no oxidizing element was introduced in the solution, cobalt nanocrystals were initially obtained from the decomposition of $Co_2(CO)_8$. Aiming to obtain nanocrystals with different sizes, reaction times were adjusted from 60 min to 120 min. These nanocrystals were oxidized at room temperature under ambient atmosphere. After room temperature oxidation, sample 2 consists of small nanoparticles with 5 nm average diameter [Fig. 1(b)]. As illustrated by XRD patterns [Fig. 2(b)], upon air oxidation at room temperature, metallic Co of the nanoparticles in sample 2 were stabilized in the divalent state (CoO).

Higher reaction times substantially promoted the particle growth. After room temperature partial oxidation, samples 3 and 4, had an average final diameter of 9.9 and 13.0 nm, respectively [Fig. 1(c)-(d)]. At earlier stages of the reaction (until 60 min), the diameter increase is attributed to the direct and uniform incorporation of monomers from $Co_2(CO)_8$ decomposition on the particles surface [18]. This is the reason for the spherical shape of nanoparticles in sample 3. When the monomer supply and, as a consequence, the reaction rate started to decrease we believe that the nanoparticles coalescence became the main growth mechanism. Such mechanism explains the slight deviation of shape from spherical in the case of sample 4, by involving a less uniform growth. The different contrast obtained by TEM between the darker core and the lighter shell of the nanoparticles after room temperature oxidation suggests their uncompleted oxidation. HRTEM images [Fig. 3(a)] indicate the formation of multiple crystalline domains within the shell of these nanoparticles as a result of the suggested growth procedure. X-ray diffraction of these samples [Fig. 2(c)-(d)] indicates that the main phase is CoO, although the presence of some $Co_3O_4$ could be anticipated [19] since $Co_3O_4$ usually exists simultaneously in CoO [20]. Similar examples, showing that the phase of the 3d-metal oxides depends on the size of the nanoparticles, have also been recently reported for Fe [21] and Mn [22], most probably indicating that cation vacancies are formed as a result of the metal oxide passivation.

Interestingly, XRD confirms that a percentage of cubic metallic Co also remains, even after particles exposure to air. Apparently, initially the oxidation reaction takes place at the oxide-cobalt interface by means of oxygen diffusion into the nanoparticle. Then, as the oxidation proceeds, the cobalt ions outward diffusion through the oxide layer start to be dominant and the reaction continues at the solvent-oxide interface [23]. The ion diffusion and thus the oxidation rate decreases with the shell thickness and for the largest particles an unreacted metallic Co nucleus remains stabilized inside the particle. Indeed, it has been previously stated that for the iron and cobalt particles some critical size exists above which full reaction with air is not possible at relatively low temperatures, and the oxidation of large enough nanoparticles results in yolk-shell nanostructures [19, 24, 25].

The 18 nm yolk-shell nanoparticles were obtained following a synthetic route identical to that used for the preparation of 13 nm core-shell nanoparticles, but maintaining the reaction solution open to the atmosphere (sample 6). To promote the complete particle oxidation, pyridine-N-oxide was introduced in the heated solution containing the nanocrystals. Thereby, in sample 5, cobalt nanocrystals were obtained from the decomposition of $Co_2(CO)_8$ at 180 ºC during 90 min and their posterior oxidation at 150 ºC during 120 min by pyridine-N-oxide .

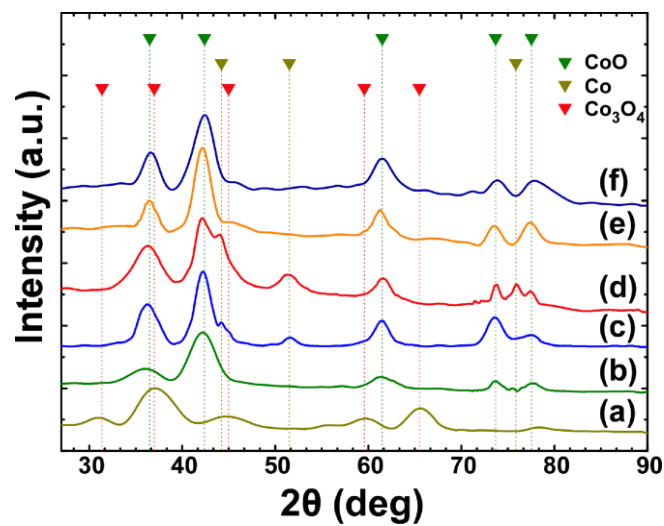

**FIG. 2.** XRD patterns of samples 1-6 (a)-(f) compared to JCPDS cards #15-0806, #48-1719 and #42-1467 corresponding to cubic Co, CoO and $Co_3O_4$.

As previously described in detail [12,19,26], the inflow of oxygen from the atmosphere during the growth stage of Co nanoparticles introduces significant variations in the observed morphology. Oxygen is taken up in such a way as to form vacant cation sites [23]. This is followed by the merging of these vacancies and their diffusion to the boundary between the core and the shell, in a process similar to the Kirkendall effect for bulk materials. The formed cores, probably consisting of cubic Co, gradually shrink by the transportation of cobalt atoms to the particle shell. The paths for such material transfer are usually one or more bridges connecting the core with the surrounding oxide shell [26], although in some cases an asymmetric distribution of the core inside the void is also observed [11].

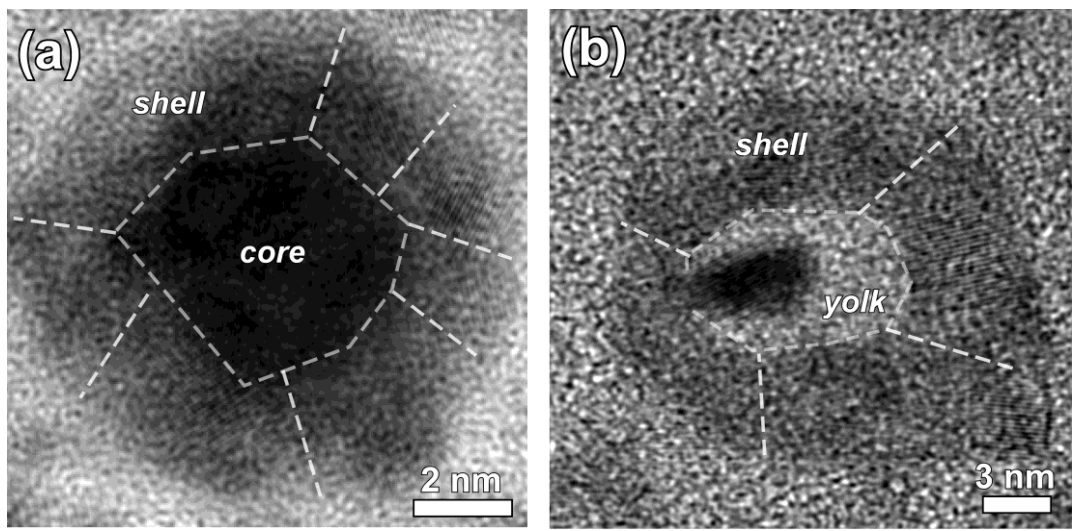

**FIG. 3.** HRTEM images illustrating nanoparticles with core-shell (a) and yolk–shell (b) configuration.

HRTEM images of the yolk-shell nanoparticles characterized in this work [Fig. 3(b)] clearly indicate an asymmetric distribution of the metal core and the appearance of the described bridges with the nanoparticle shell. The consumption of the core material is accelerated during a thermal aging stage under oxidizing conditions. In a similar case, described for Ni/NiO core-shell nanoparticles, the appearance of off-centered cores was attributed to the competitive diffusion of Ni across the interface and the growth of a void as a result of vacancies supersaturation [27]. During this sequence, the rate of Ni oxidation is reduced due to the gradually limited interfacial area and the core becomes a small sphere lying asymmetrically in the one side of the void. In our case, Fig. 1(e) for the hollow nanoparticles (sample 5) indicates the accomplishment of such process. For the same amount of Co ions, hollow and yolk-shell nanoparticles obtained by oxidation at high temperatures have obviously larger total volumes than solid Co/CoO nanocrystals. In many cases there is strong evidence that these shells have a pseudomorphic form, thus they are not in thermodynamic equilibrium [25]. This is probably an explanation for the noticed partial breakage of yolk-shell and hollow multi-crystalline particles into smaller crystallites within a period of several months.

B. GEOMETRICAL SIMULATION

In the simplest scenario, the formation of hollow oxide nanostructures by the Kirkendall effect is a two-step process. The first step involves the formation of solid metal nanoparticles. In the next step, an oxidizing element is reacted with the metal particles to yield oxide nanocrystals. Several factors determine the final geometry of the particles; mainly the self-diffusivities of the metal and oxygen ions in the oxide shell and the reaction rates at each interface. Depending on these parameters, the shell growth takes place totally or partially at the oxide/solution interface or at the metal/shell interface. When the reaction takes place at the metal/shell interface, normally due to a larger anion than cation self-diffusivity, and it proceeds to completion, solid oxide nanocrystals are obtained. In this scenario, metal-metal oxide core-shell nanoparticles are obtained when the oxidation reaction is not completed. On the other hand, when the reaction fully takes place at the solution/oxide interface, usually due to a much larger cation than anion diffusivity through the oxide layer, and the oxidation reaction proceeds to completion, hollow oxide nanoparticles are obtained. In this case, yolk-shell nanocrystals are the intermediate product obtained from an incomplete oxidation reaction. In between the two extreme cases, when the self-diffusivity is just slightly higher for the cation than for the anion and the reactivities at both interfaces are similar, core-shell nanoparticles are initially formed, but they develop into yolk-shell nanocrystals as the reaction proceeds. Hollow nanoparticles with small inner-to-outer radius ratios are the final reaction product [15]. In Fig. 4 the results of a simulation, using a simple kinetic model, of the geometry evolution of 10 nm Co nanocrystals during the oxidation reaction are represented [14]. The model assumes the ions diffusivities through the shell to be the growth limiting parameters, and considers them constant through the whole reaction. For a Co self-diffusivity much larger than that of oxygen, the shell grows totally at the shell/solution interface, yolk-shell nanoparticles are the intermediate product and hollow nanoparticles are finally obtained [Fig. 4(a)]. For equal self-diffusivities or faster anion than cation diffusion, no vacancies accumulate at the core/shell interface and thus core-shell nanoparticles are the intermediate oxidation product and solid oxide nanoparticles are finally obtained [Fig. 4(c)]. However, when the cation self-diffusivity is just slightly higher than that of the anion, the shell growth takes place partially at the solution/shell interface and partially at the core/shell interface. Notice how, in this scenario, the outer shell diameter monotonically increases with the reaction time, while the core and inner shell diameter decrease. At the initial reaction stage, the inner shell diameter follows the decrease of the core diameter, thus core-shell nanoparticles are being formed. At a certain reaction time, a difference appears between the core and the inner shell diameters, thus the formation of yolk-shell nanocrystals becomes evident. In this case, the final oxide nanoparticles are hollow, but have relatively small inner shell diameters. In the model, the evolution from a core-shell to yolk-shell structure is obtained due to the increasingly higher weight of the self-diffusivities difference in the inner-to-outer shell growth ratio. Additionally, again for the same amount of reacting metal ions, the smaller the core size, the higher the rate of core shrinkage is acheived. At the same time, the increasing concentration of vacancies in a decreasing interface area, makes the separation between the core and the shell more evident, the smaller the core size becomes.

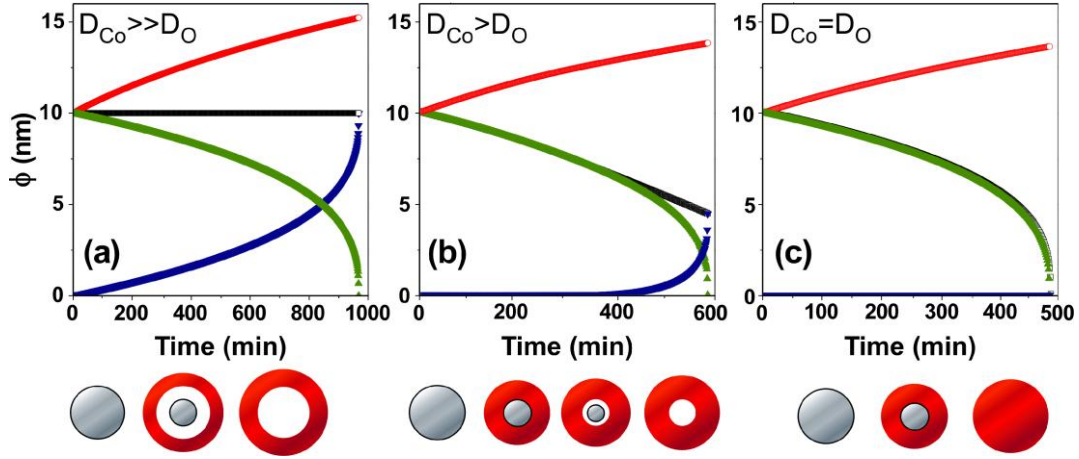

**FIG. 4.** Geometry evolution during the oxidation of 10 nm Co nanoparticles considering the following cobalt vs. oxygen diffusivity ratios: (a) $D_{Co}/D_O = 1000$; (b) $D_{Co}/D_O = 10$; (c) $D_{Co}/D_O = 1$. ■: Inner shell diameter (φi); ○: outer shell diameter (φo); ▲: core diameter (φc); ▼: φi-φc.

## C. MAGNETIC CHARACTERIZATION

As previously described, by keeping the reactant quantities fixed, the reaction duration and the presence of oxidants mainly determine the final size and the oxidation extent in each sample. Consequently, the magnetic properties and the correlated nanoscale effects in these particles can be adjusted.

Hysteresis loops of the samples were recorded in the range 10-300 K. The oxidized samples 1 and 2 present net antiferromagnetism at room temperature [Fig. 5(a)-(b)]. Upon cooling, a weak FM contribution appears in 5 nm CoO nanoparticles. Such effect is attributed to the stabilization of uncompensated magnetic moments on the surface of the nanoparticles that is enhanced below the Néel temperature of CoO at around 290 K [28]. In this regard, experimental evidences of finite-size effects and uncompensated magnetization in AFM CoO have been previously reported in thin films [29], with saturation values closely matching our results for the range of thicknesses discussed here. It is intuitively suggested that magnetization values would increase with the increase in the number of atoms at the interface, as for the case of hollow CoO nanoparticles. This is observed in Fig. 5(b-c). In addition, samples with hollow morphology are composed of polycrystals (Fig. 3) with smaller grain sizes than the spherical particles [12], resulting in a spread of the anisotropy axis across the surface and a higher density of uncompensated spins [30]. The enhancement of the surface anisotropy resulting from the additional inner spin disordered surface of the particles with hollow morphology is noticeable as it leads to an increase in the blocking temperature and the appearance of hysteresis above 300 K [Fig. 5(c)]. Similar effects have been recently demonstrated also in hollow $NiFe_2O_4$ particles [25]. On the contrary, $Co_3O_4$ nanoparticles [Fig. 5(a)] remain antiferromagnetic despite their higher surface-to-volume ratio since Néel relaxation of $Co_3O_4$ for this diameter should occur at least under 33 K [31].

For the core-shell cases, shown in Fig. 6(a) and 6(c), the room temperature saturation magnetization values are found to be 25 and 9 $Am^2/Kg$, respectively. Assuming the negligible role of the antiferromagnetic shell in the determination of these values and estimating the percentage of organic compounds by thermogravimetric analysis (not shown, but approaching 50 % for both samples), it was possible to calculate the non-oxidized cobalt content of each sample. Core-shell nanoparticles of sample 3 were found to contain around 40 %wt of cobalt compared to 15 %wt of sample 4. The corresponding core-to-shell ratio in terms of volume implies a core diameter of 5 nm in both samples and an oxide shell thickness of 2.5 and 4 nm in sample 3 and 4, verifying the estimations from TEM analysis. Thicker shells and smaller core volumes are found for sample 6.

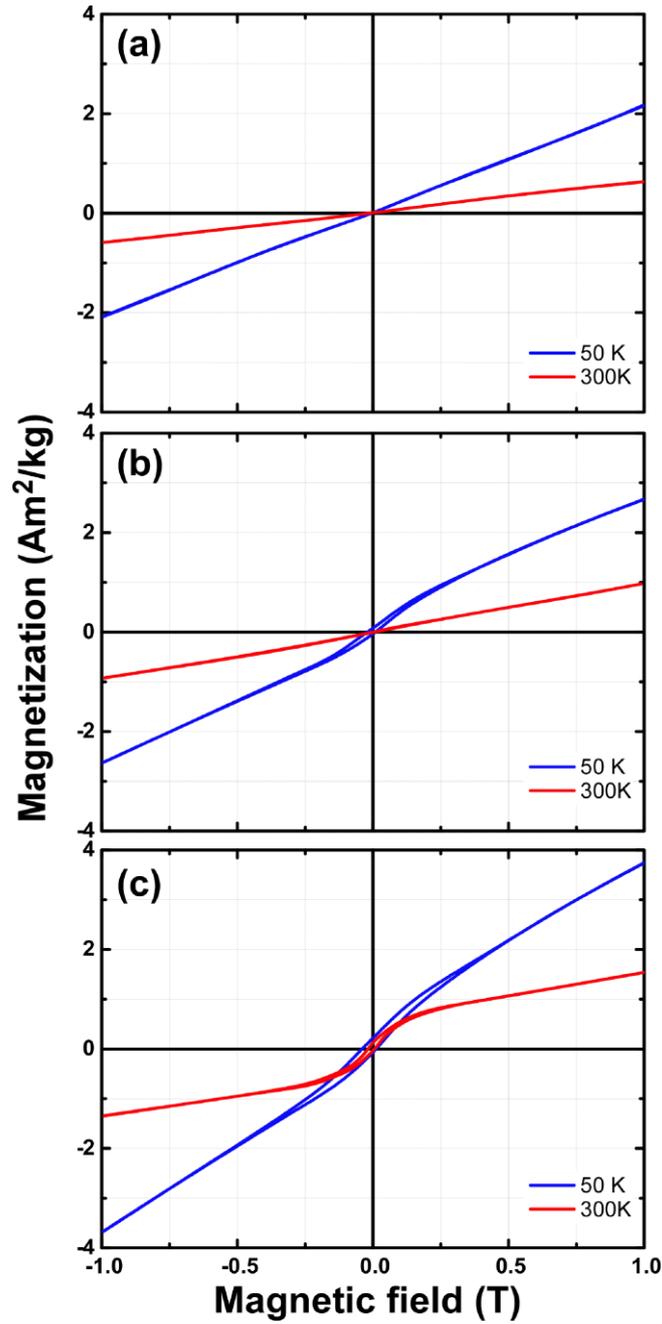

**FIG. 5.** Hysteresis loops of samples 1 (a), 2 (b) and 5 (c) at 50 and 300 K.

Furthermore, the absence of saturation at lower temperatures [Fig. 6(a),(c) and (e)] could be attributed to the increasing alignment of the shell spins to the core magnetization, at high applied fields [32]. Magnetization versus temperature M(T) measurements give further insight into the magnetism of such systems. There are two important features of the ZFC curve in the low temperature region. The part where magnetization is constantly close to zero is the result of the random orientation of particles magnetization. The transition from zero magnetization to the rapidly increasing part is described as the exchange-bias onset temperature [33]. After this point the FM/AFM coupling begins to gradually disappear. Experimentally, we also find a sudden rise of ZFC-FC magnetization below 33 K, attributed to traces of $Co_3O_4$ [24]. The $Co_3O_4$ shows two spinel sites, the $Co^{2+}$ ions occupying the tetrahedral sites and the $Co^{3+}$ ions the octahedral positions. In passing its Néel temperature (33 K) the octahedral sites properties remain unchanged, being the $Co^{3+}$ diamagnetic, while the $Co^{2+}$ ions form an AFM magnetic sublattice, and thus explaining the upturn in the M(T) curves at low temperatures. Nevertheless, formation of $CoCO_3$ as an intermediate product during decomposition is another possibility taking into account its low Néel temperature (18 K).

Noteworthy, the corresponding ZFC-FC curves of samples 3 and 4 are indicative of the presence of a FM system at temperatures as high as 300 K. The blocking temperature for superparamagnetism lies over room temperature although a much lower value should be expected for Co nanoparticles with a diameter equal to the core dimension (5 nm). We surmise the coupling of Co core with the oxide shell is responsible for a huge increase of blocking temperature, which was previously reported to reach almost two orders of magnitude [2]. The strong pinning of Co magnetic moments is more obvious at the part of ZFC branch where magnetization gradually aligns to the external field. While in sample 3 magnetization increases monotonically overcoming 50 K, in sample 4, the ZFC curve presents a delay in the temperature of rise (150 K) as well as a step (between 230 - 270 K) before it meets the FC branch. Similar arguments apply for sample 6. This is indicative of the AFM shell thickness dependency of exchange coupling. Specifically, the stepwise approaching of maximum magnetization value could be attributed to the existence of non-compensated interfacial spins pinned to the AFM material requiring higher thermal energy to decouple. The fact that the Néel temperature of CoO with a volume corresponding to the 4 nm shell is also expected to occur in this range provides a better explanation of the effect through the loss of AFM ordering. It should be mentioned that due to finite size effects, the Néel temperature of bulk CoO (~ 290 K) may diminish to 150 K for a 2 nm-thick CoO shell [34], thus a lower value is estimated for sample 3.

As expected, the interaction between Co and the oxidized shell results in the appearance of a shift of the hysteresis loop center to negative fields at lower temperatures. Actually, comparison of samples 3 and 4 shows the exchange bias field increases monotonically with decreasing temperature (Fig. 7). Both samples follow the same trend but the intensity of exchange interaction appears to be higher for sample 4, taking the maximum value of 2.1 kOe at 10 K, while the corresponding value for sample 3 is 1.6 kOe.

The intensity of the interaction between the AFM shell and the FM core is proportional to the dimensions of the oxide coating and to the core volume [21]. Therefore, the enhancement of exchange bias, for about 30 % comparing samples 3 and 4, is attributed to the thickening of the shell since the cobalt core diameter remains the same (5 nm). For instance, the anisotropy energy of a CoO shell is proportional to the term $(R^3-r^3)$, where R and r is the particle and the core radius [35]. Accordingly, the intensity of exchange interaction may increase significantly as shell thickness grows, keeping the core radius constant, till anisotropy energy overcomes the effective Zeeman energy of the FM core. In this case, anisotropy energy is expected to be doubled for sample 4, thus explaining the rise of exchange bias.

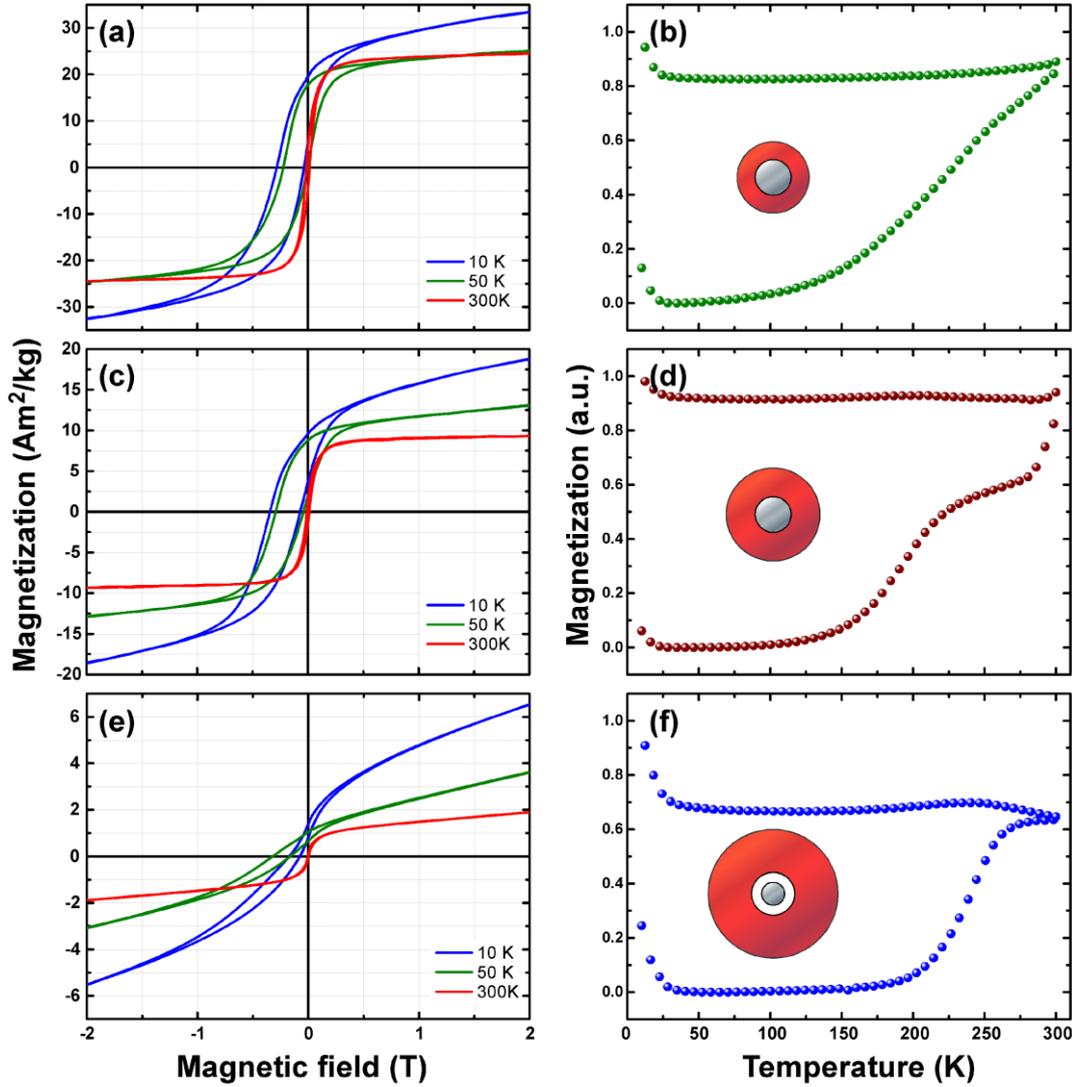

**FIG. 6.** Hysteresis loops and ZFC-FC curves measured at the temperature range 10-300 K corresponding to samples 3 (a,b), 4 (c,d) and 6 (e,f).

The results concerning yolk-shell Co/CoO nanoparticles [Fig. 6(e) and 6(f)] are more surprising since exchange bias appear to take higher values despite significantly smaller dimensions of the FM/AFM interface, limited in the bridges between Co and the shell. Furthermore, the fact that core diameter is only 3 nm indicates the presence of a very strong coupling effect stabilizing a FM behavior up to about 290 K, as the corresponding ZFC-FC curves illustrate. Trying to explain such unexpected effect, two points should be noted. The first is the large shell-to-core volume ratio and the non-symmetric arrangement that signifies a higher orientation degree for the frozen interfacial spins (uniaxial anisotropy) compared to the radial topology (multiaxial) case for the core/shell samples. This is consistent with the hypothesis that only a small fraction of interfacial spins control the exchange bias [30,36]. A second indirect consideration is the low saturation magnetization of the system due to the high Co-oxide content compared to samples 3 and 4, which results in the amplification of exchange bias [33]. However, the reduction of exchange bias when the temperature drops below 50 K is not clear at present. We tentatively attribute this effect to a slight canting of the $Co^{2+}$ spins at the tetrahedral sites within the $Co_3O_4$. Some support of this view is suggested by the magnetization measurements on sample at low temperatures [Fig. 6(e)]. The asymmetric "humming bird-like" shape of the hysteresis loop at 10 K is generally attributed to a different magnetization reversal mechanism occurring in each branch rather than the uniform magnetization. Whereas a magnetic single-domain nanostructure with uniaxial magnetic anisotropy follows Stoner-Wohlfarth switching behavior, the magnetic hollow particles may reverse in a rather inhomogeneous way, as shown by Goll et al. [37]. Thus, the "humming bird-like" hysteresis is an indication of exchange coupling between two magnetic phases, an interfacial hard phase (Co/CoO) and a soft phase ($Co_3O_4$). It should be noted the anisotropy constant for AFM CoO amounts about $10^8$ erg/cm$^3$ [2], while estimations for $Co_3O_4$ drop to $10^4$ erg/cm$^3$ [31]. None of these phases were clearly detected by X-ray diffraction, but the presence of $Co_3O_4$ is well evidenced in the ZFC-FC measurements. Accordingly, this asymmetry is only observed at temperatures below the Néel temperature for $Co_3O_4$ (it does not show at 50 K).

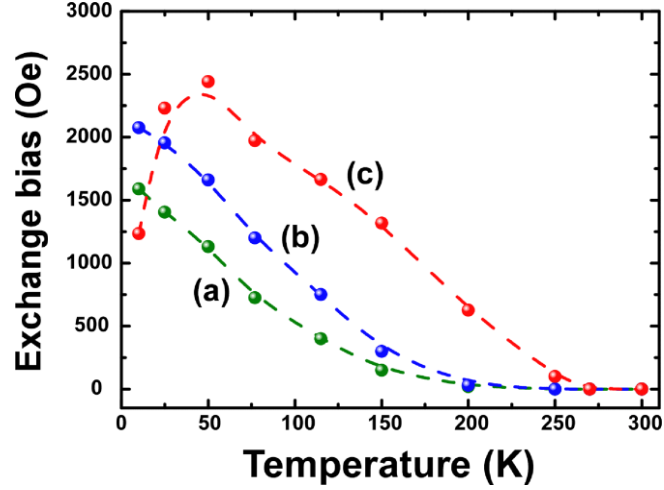

**FIG. 7.** Temperature dependence of the exchange bias field for samples 3 (a), 4 (b) and 6 (c).

### D. MAGNETIC SIMULATION

In order to gain some insight on the peculiarities of the experimental magnetic behavior presented in the previous section, we have conducted a Monte Carlo simulation study of individual nanoparticles with morphologies similar to the ones of the samples studied experimentally. The simulations use the standard Metropolis algorithm and are based on a Heisenberg model for classical spins in a sc lattice with the following Hamiltonian:

$$H/k_B = -\sum_{\langle i,j \rangle} J_{ij} \left( \vec{S}_i \cdot \vec{S}_j \right) - \sum_i \vec{h} \cdot \vec{S}_i + E_{anis} \quad (1)$$

$$E_{anis} = -k_S \sum_{i \in S} \sum_{j \in nn} \left( \vec{S}_i \cdot \hat{r}_{ij} \right)^2 - k_C \sum_{i \in C} \left( \vec{S}_i \cdot n_i \right)^2 \quad (2)$$

which includes the nn exchange interactions, the Zeeman energy with h= μH/$k_B$ (H is the magnetic field and μ the magnetic moment of the magnetic ion), and the magnetocrystalline anisotropy energy $E_{anis}$ explicitly takes into account the distinct anisotropies: Néel type for surface spins ($K_S$) having coordination reduced with respect to bulk and uniaxial along the field direction for the rest ($K_C$).

We start by analyzing the results for particles having core (FM)/shell (AFM) geometry with $J_{ij}$ = 10 K and $K_C$= 0.22 K/spin for core spins and $J_{ij}$ = -0.23 K and $K_S$ = 40 K/spin for shell spins as is the case of samples 3 and 4.

In Fig. 8, we show the simulated hysteresis loop after a FC process of an individual particle with the same characteristics as those of sample 3. First, we notice that the qualitative shape of the experimental loops, with high irreversibility fields and linear high field susceptibility, is reproduced only when the particle has crystallites at the surface. The horizontal loop shift is also well reproduced by the simulation results (blue circles in Fig. 8). The high field behavior of the particle is dominated by the shell spins (red circles in Fig. 8) and, in particular, the surface shell spins are responsible for the superimposed linear component in the hysteresis loop, as can be clearly seen when comparing with the loop for a particle with a single crystal shell (see Inset in Fig. 8), which has a more squared shape and is completed saturated in the same field range. The origin of the horizontal loop shift can be traced back to the contribution of interfacial spins at the shell (green triangles in the lowest inset of Fig. 8), some of which remain pinned during magnetization reversal of the core. All these features are observed only for high enough surface Néel anisotropy compared to bulk values. Furthermore, the relevance of the polycrystalline nature of the CoO shell on the peculiar magnetic behavior displayed by the core/shell particles is corroborated by the upper inset in Fig. 8, which shows that the hysteresis loop of a particle with identical dimensions but with a single crystalline shell does not show the high field magnetic response present in the particle with a polycrystalline shell.

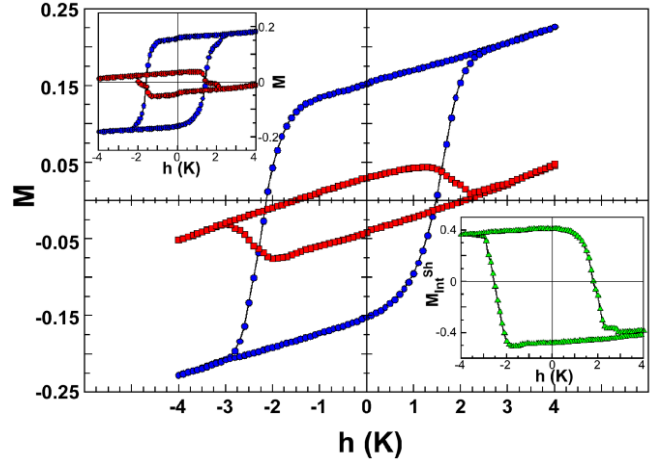

**FIG. 8.** Hysteresis loop for a core/shell particle with the same dimensions as sample 3 (blue squares). The contribution of the shell spins is shown in red circles. The lowest inset shows the contribution of interfacial shell spins ($M_{int}^{Sh}$) to the hysteresis loop while the upper inset displays the total (blue circles) and shell contributions to the hysteresis loop of a core/shell particle with identical size but with single crystalline shell.

We have studied also the role played by the shell thickness on the magnetic behavior by computing the hysteresis loop for a particle with a thicker shell and almost the same core size as is the case of particles of sample 4. The results, shown in Fig. 9, agree with qualitative change in the loop shape and are in good agreement with the one observed experimentally. The loop asymmetry present in the experimental result [Fig. 6(b)] is also observed in the result of the simulation and can be understood by realizing that the reversal mechanisms in the increasing and decreasing field branches are not the same. The snapshots of the magnetic configurations displayed in the insets of Fig. 9 show that this indeed is the case. The apparent vertical displacement of both the experimental and simulated loops (that is not observed for the thinner shell particle), can be understood by looking at snapshots taken at the remanence points (not shown), that indicate different remanent states of the particle core induced by the strongest exchange coupling to the different shell crystallites.

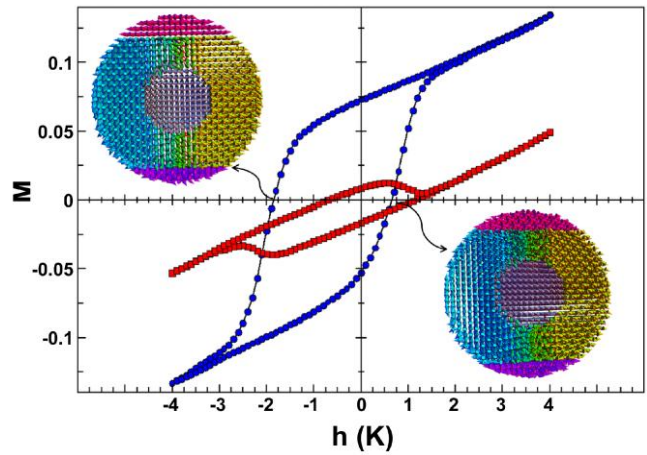

**FIG. 9.** Hysteresis loop for a core/shell particle with the same dimensions as sample 4 (blue circles). Snapshots show a slice of the core/shell particle through a plane parallel to the applied magnetic field. Crystallites at the shell are distinguished by different colors; spins at the yolk have been colored in gray. Units axones

Moreover, the loop shift for the particle with thicker shell is higher than the one for the thinner shell particle in the whole range of simulated temperatures (see Fig. 10 for the thermal dependence of the loop shift), similar to what was observed experimentally in Fig. 7.

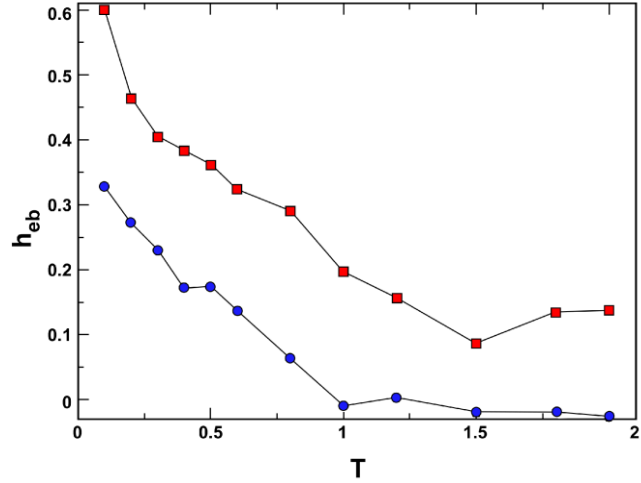

**FIG. 10.** Thermal dependence of the exchange bias field for the core/shell particles of Fig. 8 (sample 3, blue circles) and 9 (sample 4, red squares) that have the same core dimensions but different shell thickness. Units axones

Finally, we focus on the behavior of yolk/shell particles, which have been modeled by displacing the FM core center of the core/shell particles down by a distance $d_c$. An example of the simulated hysteresis loops for a particle with $d_c = 10$ is shown in Fig. 11. As can be seen, the loop shift for the yolk/shell particle is higher than that for the core/shell particle with the same dimensions. This demonstrates that in fact only a small fraction of interfacial shell spins controls the magnitude of the loop shift, as it is also confirmed by comparing the interfacial spin contributions in both cases (green diamonds in Figs. 8 and 9). The coercive field of the yolk/shell particle is lower than that of a core/shell one and has a more elongated shape, in agreement with experimental results in Fig. 6. The lowest magnetization of the simulated yolk/shell nanoparticle is also in agreement with the experimental observation. These last observations can be traced back to the different reversal mechanisms in both field branches and the high degree of disorder present at the particle surfaces, as indicated by the representative snapshots of the magnetic configurations near the coercive fields shown in the insets of Fig. 11.

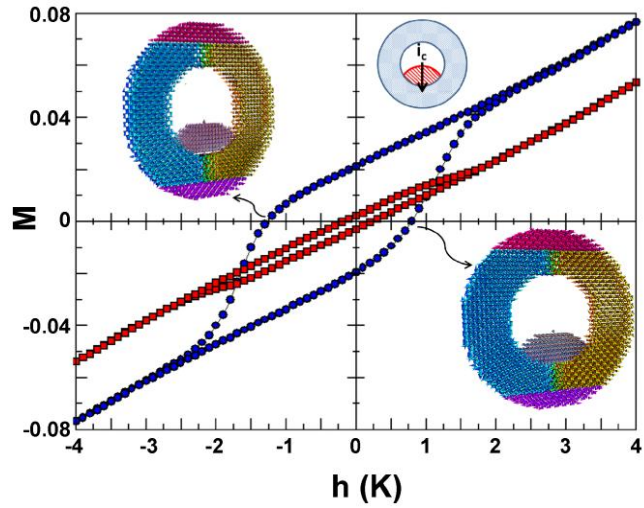

**FIG. 11.** Hysteresis loop for a particle with a yolk structure similar to the one in sample 6 (blue squares). The yolk has been built by displacing the core center down by $d_c$ unit cells (see the schematic drawing in the central inset). The contribution of the shell spins is shown in red circles. The left and right insets display the magnetic configurations near the coercive field points. Snapshots show a slice of the yolk/shell particle through a plane parallel to the applied magnetic field. Crystallites at the shell are distinguished by different colors; spins at the yolk have been colored in gray. Units-axonas-y

## IV. CONCLUSIONS

The conditions affecting the composition, the morphology and the magnetism of Co/CoO nanoparticles prepared by $Co_2(CO)_8$ decomposition were studied. When the synthesis proceeds under inert environment, partially oxidized Co core/Co-oxide shell nanoparticles are obtained after exposure to air. As a result, exchange coupling, with intensity proportional to the shell thickness, determines the magnetic properties of these samples. Oxygen presence during the reaction favors the formation of hollow or yolk-shell particles through the nanoscale Kirkendall mechanism. In this case, uncompensated spins are efficient to enhance the expected anisotropy for an oxide or to induce intense exchange effects. Experimental results are totally consistent to the kinetic model simulation used for the growth of nanoparticles and the Monte Carlo numerical studies of their corresponding magnetic behavior.

**ACKNOWLEDGMENTS**

C. M. Boubeta and A. Cabot grant financial support through the Ramón y Cajal program. A. Cabot also acknowledges financial support through the Spanish MICINN Projects MAT2008-05779, MAT2008-03400-E/MAT and ENE2008-03277-E/CON. O. Iglesias thanks financial support from Spanish MICINN through project MAT2009-0866 and Generalitat de Catalunya through project 2009SGR876 and acknowledges CESCA and CEPBA under coordination of $C^4$ for supercomputer facilities.